\documentclass[%
 reprint,
groupedaddress,
footinbib,
bibnotes,
amsmath,amssymb,
aps,
nolongbibliography
]{revtex4-1}

\usepackage{graphicx}
\usepackage{dcolumn}
\usepackage{bm}
\usepackage{lipsum}

\usepackage{xcolor}

\newcommand{\br}{\mathbf{r}}
\newcommand{\bfo}{\mathbf{f}}

\newcommand{\be}{\mathbf{e}}
\newcommand{\bn}{\mathbf{n}}

\newcommand{\cC}{\mathcal{C}}
\newcommand{\cE}{\mathcal{E}}

\newcommand{\id}{\mathrm{d}} 
\renewcommand{\d}[2]{\frac{\id #1}{\id #2}} 
\newcommand{\dd}[2]{\frac{\id^2 #1}{\id #2^2}} 

\newcommand{\pd}[2]{\frac{\partial #1}{\partial #2}} 
\newcommand{\pdd}[2]{\frac{\partial^2 #1}{\partial #2^2}} 
 
\begin{document}

\preprint{APS/123-QED}


\title{\textit{Gyrophilia}: Harnessing Centrifugal and Euler Forces for \\ Tunable Buckling of a Rotating \textit{Elastica}}

\author{Eduardo Gutierrez-Prieto}
\author{Michael Gomez}%
\author{Pedro M.~Reis}
\email{pedro.reis@epfl.ch}
\affiliation{Flexible Structures Laboratory, Institute of Mechanical Engineering, \'{E}cole Polytechnique F\'{e}d\'{e}rale de Lausanne (EPFL), 1015 Lausanne, Switzerland}

\date{\today}

\begin{abstract}
We investigate the geometrically nonlinear deformation and buckling of a slender elastic beam subject to time-dependent `fictitious' (non-inertial) forces arising from unsteady rotation. Using a rotary apparatus that accurately imposes an angular acceleration around a fixed axis, we demonstrate that centrifugal and Euler forces can be combined to produce tunable structural deformation. Specifically, using an imposed acceleration ramp, the buckling onset of a cantilevered beam can be precisely tuned and its deformation direction selected. In a second configuration, a pre-arched beam can be made to snap, on demand, between its two stable states. We also formulate a theoretical model rooted in Euler's \textit{elastica} that rationalizes the problem and provides predictions in excellent quantitative agreement with the experimental data. Our findings demonstrate an innovative approach to programmable actuation of slender rotating structures.
\end{abstract}

\maketitle

Nearly every modern machine involves rotary elements~\cite{shigley_mechanical_1985} (\textit{e.g.,} shafts, wheels, bearings, fans, and turbines), which are so ubiquitous that they often go unnoticed. Since the pioneering studies on rotating shafts by Rankine~\cite{Rankine_on_the_centrifugal}, F{\"o}ppl~\cite{foppl1895}, and Jeffcott~\cite{jeffcott1919} over a century ago, predictive modeling has become essential in designing and analyzing rotating machinery. \emph{Rotordynamics}~\cite{rao_history_2011,childs_turbomachinery_1993,dimarogonas2013analytical,muszynska_rotordynamics_2005,genta2005dynamics} has since evolved into a mature field with the primary focus of understanding the vibratory dynamics of rotating structures, to prevent large-amplitude motions that may cause catastrophic failure.  Representative examples from the vast literature on rotating structures include improving the operational range and efficiency of jet-engine turbines~\cite{srinivasan_flutter_1997},  turbo-compressors~\cite{verstraete_multidisciplinary_2010}, hydraulic machines~\cite{gomes_pereira_prediction_2022}, centrifugal microfluidics~\cite{gorkin_centrifugal_2010,kong_lab---cd_2016,bhamla2017hand}, and space structures~\cite{moravec_non-synchronous_1977,wu_heliogyro_2018,gardsback2009deployment}. 

When formulating Newton's equations of motion in a rotating (non-inertial) frame of reference (FoR), three `fictitious' body forces appear to act on a rotating body~\cite{tipler_physics_2002}: (i) the \textit{centrifugal} force (proportional to the square of the angular velocity); (ii) the \textit{Coriolis} force (resulting from FoR-body relative motion); and (iii) the \textit{Euler} force (opposing angular acceleration). Prior research has investigated rotation-induced instabilities in slender structural elements including rods~\cite{mostaghel_buckling_1973,white_buckling_1979,aganovic_stability_2001,richard_buckling_2018}, plates~\cite{nowinski1964,chen2007,chen2011,delapierre_wrinkling_2018,coman_wrinkling_2023}, and shells~\cite{lam_analysis_1995,li_rotating_2005}. These studies primarily considered the centrifugal forces caused by constant angular velocities, sometimes accompanied by Coriolis forces~\cite{sane_antennal_2007,sreenivasamurthy_coriolis_1981,banerjee_dynamic_2014}, but rarely taking the effects of Euler forces into account. As an exception, motivated by the `spin-up' of disk drives, the stress distribution and wrinkling of rotating elastic disks have been quantified~\cite{tang_note_1970,sader_shear-induced_2019}.

\begin{figure}[b!]
  \centering
  \includegraphics[width=\columnwidth]{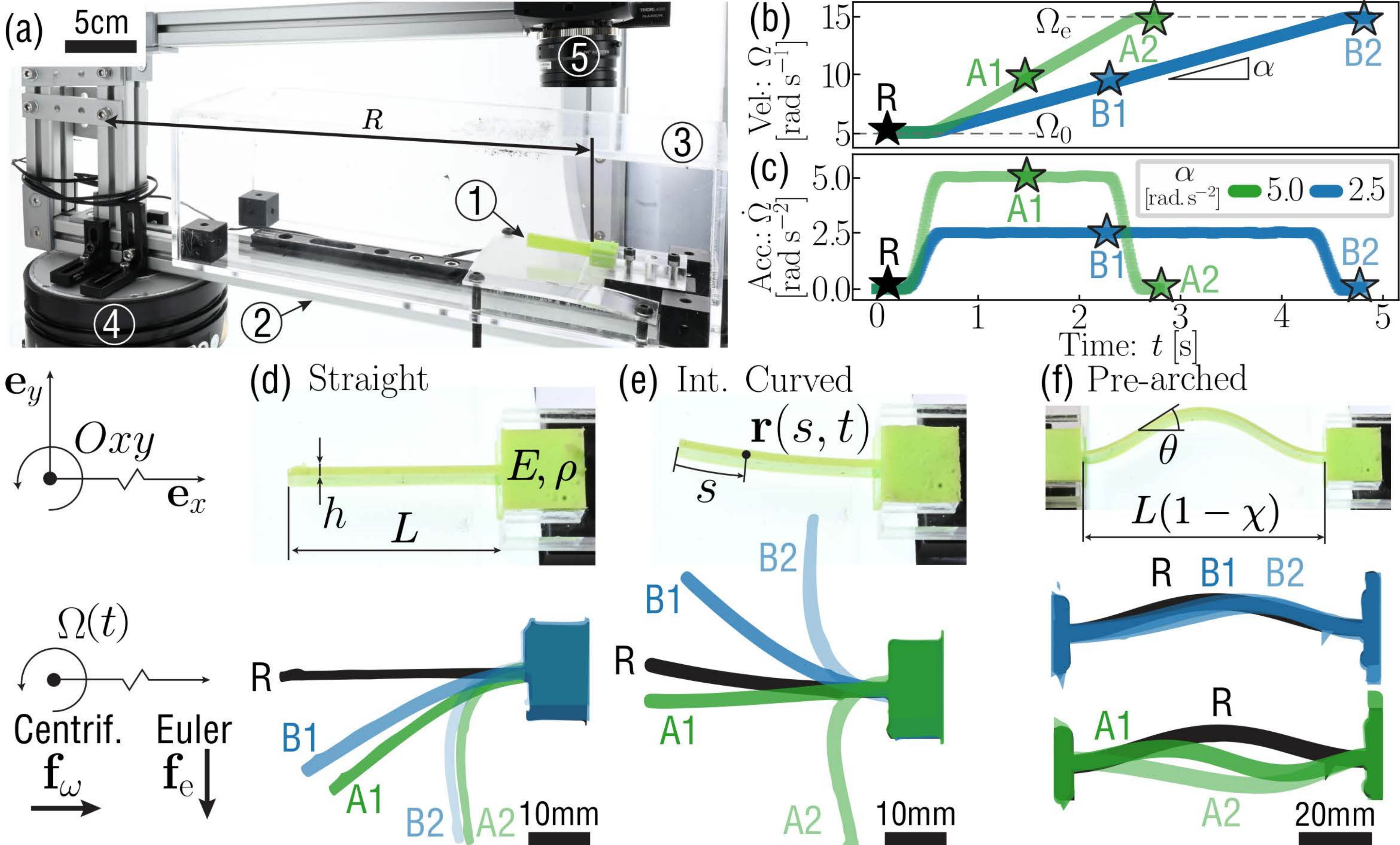}
  \caption{(a) A beam (1) is mounted on a rigid arm (2) inside an acrylic box (3). A torque motor (4) rotates the arm and a camera (5). (b) Representative time-series of the imposed angular velocity, $\Omega(t)$, and (c) angular acceleration, $\dot{\Omega}(t)$, with $\alpha{=}\{5,\, 2.5\}\:\mathrm{rad}/\mathrm{s}^{2}$ (green/blue curves; see legend) and $(\Omega_0,\,\Omega_e){=}(5,\,15)\:\mathrm{rad}/\mathrm{s}$. Bottom panels: undeformed (top row) and deformed (lower row) specimens for (d) straight and (e) naturally curved cantilevers; and (f) pre-arched (double-clamped) beam. Deformed configurations are taken at the instances labeled in panels (b,\,c). See also S.M. Video 1-3~\cite{endnoteSuppInfo}.}
  \label{fig1}
\end{figure}

Here, we perform experiments on unsteadily rotating elastic beams placed eccentrically about a fixed axis (Fig.~\ref{fig1}a--c and S.M. Video 1~\cite{endnoteSuppInfo}). We also conduct simulations of a dynamic model based on Euler's \emph{elastica}~\cite{audoly2010}, specialized to a rotating FoR. Two configurations are examined: cantilevered beams (clamped-free ends) and beams pre-buckled into a bistable arch (double-clamped). In both cases, the loading arises primarily from centrifugal and Euler forces. For cantilevered beams, beyond a critical angular velocity, the centrifugal force (along $+\be_{x}$; see Fig.~\ref{fig1} bottom-left) triggers a buckling instability (Fig.~\ref{fig1}d). Simultaneously, the Euler force (along $-\be_{y}$) acts as a symmetry-breaking `imperfection' that selects the buckling direction, potentially opposing the beam's natural curvature (Fig.~\ref{fig1}e). For arched beams, these forces switch roles: the Euler force drives snap-through buckling while the centrifugal force modulates the asymmetry of the arched shape and the acceleration threshold for instability (Fig.~\ref{fig1}f). Following an approach we term \emph{gyrophilia}, our study highlights how unsteady rotational loads can be leveraged for function in a new class of tunable mechanisms.

The beam specimens are cast from vinyl polysiloxane (Elite Double, Zhermack): VPS32 (Young's modulus $E {=} 1.164\,\mathrm{MPa}$, density $\rho {=} 1170\,\mathrm{kg}/\mathrm{m}^{3}$) for the cantilevered beams and VPS22 ($E {=} 863\,\mathrm{kPa}$, $\rho {=} 1190\,\mathrm{kg}/\mathrm{m}^{3}$) for the pre-arched beams ~\cite{grandgeorge_elastic_2022,leroy-calatayud_tapered_2022}. The casting is achieved using laser-cut acrylic molds to yield uniform, rectangular beams of width $b {=} 10\:\mathrm{mm}$, thickness $h{\in}[1.8,\,2.3]\,\mathrm{mm}$, length $L{\in}[40,\,100]\,\mathrm{mm}$ and (constant) natural curvature $\kappa_0{\in}{\pm}[0,\,5]\,\mathrm{m}^{-1}$. The latter range is only for the cantilevered beams; the pre-arched beams have $\kappa_0{=}0$. Each specimen is mounted onto a rigid arm attached to a high-torque motor (ETEL RTMBi140-030). An encoder records the angular position of the system at $20\,\mathrm{kHz}$, from which the angular velocity, $\Omega(t)$, and angular acceleration, $\dot{\Omega}=\text{d}\Omega/\text{d}t$, are computed (Fig.~\ref{fig1}b,\,c). See the S.M.~\cite{endnoteSuppInfo} for additional details on the rotation protocol. The beams are clamped in the radial direction $\be_{x}$, with the outer end at $R{\in}[350,\,700]\,\mathrm{mm}$ from the center of rotation (Fig.~\ref{fig1}a). The inner end is either  free (cantilevered beams; Figs.~\ref{fig1}d,\,e) or clamped at a distance $L(1-\chi)$ radially inwards from the outer end, where $\chi\in(0,1)$ (pre-arched beams; Fig.~\ref{fig1}f). A digital camera (Mikrotron Eosens mini1, 100-550 fps) mounted onto the rotating FoR records the deformed beams, whose centerlines are extracted via image processing. A transparent box protects each specimen against air drag. The experimental system (motor, encoder, camera) is automated, enabling a systematic exploration of parameters.

\begin{figure}[t]
  \centering
  \includegraphics[width=\linewidth]{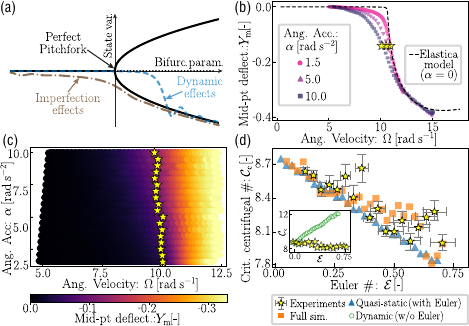}
  \caption{Buckling of naturally straight, cantilevered beams. (a) Typical response diagrams near a supercritical pitchfork bifurcation. A `perfect' pitchfork (black curves) is not expected for our system due to either dynamic effects (dashed curve) or symmetry-breaking imperfections (dashed-dotted curve).
  (b) Normalized midpoint deflection, $Y_\mathrm{m} = y_\mathrm{m}/L$, obtained experimentally as the angular velocity, $\Omega$, is ramped with acceleration $\alpha{=} \{1.5,\,5.0,\,10.0\}\:\mathrm{rad}/\mathrm{s}^{2}$ (symbols; see legend). The dashed curve represents the equilibrium solution of Eqs.~\eqref{eqn:elasticaforcedim}--\eqref{eqn:elasticamomentdim}. (c) `heatmap' of $Y_\mathrm{m}$ (see colorbar) vs.~$\Omega$ and $\alpha$. Here and in (b) stars indicate the onset of $|Y_\mathrm{m}|\geq0.15$. (d) Critical centrifugal number $\mathcal{C}_c$ (onset of $|Y_\mathrm{m}|\geq0.15$) vs.~Euler number, $\mathcal{E}$, for experiments and simulations (see legend).}
  \label{fig2}
\end{figure}

First, we perform a series of experiments on rotating, straight ($\kappa_0 = 0$), cantilevered beams (Fig.~\ref{fig1}d). The centrifugal force acts along the axis of the undeformed beam ($+\be_{x}$), exerting a compressive distributed load. Above a critical angular velocity, $\Omega_{\mathrm{c}}$, the beam buckles, causing it to bend abruptly toward $-\be_{y}$, the direction of the Euler force (Fig.~\ref{fig1}d, bottom). When $\Omega$ is varied quasi-statically ($\dot{\Omega}\approx 0$), the scenario is analogous to the buckling of a vertical cantilever under self-weight~\cite{greenhill1881,thompson1971,white_buckling_1979}. This gravity-induced buckling is described by a supercritical pitchfork bifurcation~\cite{wang1986,virgin2004}, as sketched in Fig.~\ref{fig2}(a) (solid/dotted curves). We anticipate that the buckling in our rotating system also corresponds to a supercritical pitchfork bifurcation, but only when $\dot{\Omega}\approx 0$. 

For the cantilevered beam rotated with the angular acceleration $\dot{\Omega}$ reaching the plateau value of $\alpha$ (Fig.~\ref{fig1}b,\,c), there are, \emph{a priori}, two possible, opposing effects: (I) Due to dynamic effects, the buckling onset may occur at a \emph{higher}  $\Omega$ compared to the quasi-static scenario (dashed curve, Fig.~\ref{fig2}a). Such delayed bifurcations are typically associated with dynamical systems involving a bifurcation parameter varied at finite rate~\cite{su2001,tredicce2004,liu2021}, as is the case with $\Omega(t)$ here. (II) Alternatively, buckling may occur at a \emph{lower}  $\Omega$ (dashed-dotted curve, Fig.~\ref{fig2}a) due to the asymmetry (or `imperfection') introduced by the Euler force, similar to how a transverse force (or natural curvature) lowers the buckling onset of a column under self-weight~\cite{virgin2004}. To discern whether effect (I) or (II) dominates, we vary $\alpha$ in the experiments while fixing all other parameters. In Fig.~\ref{fig2}(b), we plot data for the beam's normalized midpoint deflection, $Y_\mathrm{m}=y_\mathrm{m}/L$, versus the instantaneous angular velocity, $\Omega$, for three different accelerations, $\alpha {\in} \lbrace1.5,5.0,10.0\rbrace\:\mathrm{rad}/\mathrm{s}^{2}$. We find that a higher $\alpha$ decreases the buckling onset and smooths the perfect pitchfork shape, as expected for scenario (II). This behavior is further evident in Fig.~\ref{fig2}(c), which shows a `heatmap' of $Y_\mathrm{m}$ versus both $\Omega$ and $\alpha$. These results demonstrate that effect (II) prevails: despite the dynamic nature of the loading, the buckling instability is dominated by the symmetry-breaking effect of the Euler force.

Next, we formulate a geometrically nonlinear model for the elastic deformation of a beam undergoing unsteady rotation and, thus, loaded by fictitious forces. We adopt the \textit{elastica} framework~\cite{audoly2010}, albeit in a rotating FoR, to describe both a cantilevered beam (Fig.~\ref{fig1}d,\,e; experimental results above) and a pre-arched beam (Fig.~\ref{fig1}f; discussed below). These two configurations are distinguished by the respective boundary conditions (BCs) applied to their inner end (free or clamped, respectively). We define Cartesian coordinates in the rotating FoR with unit vectors $\lbrace\be_x, \be_y, \be_z\rbrace$, such that the beam's outer end lies on the $x$-axis; see Fig.~\ref{fig1}. Since the beam's dimensions satisfy $h \ll b \ll L$, we assume that the beam undergoes planar ($x$-$y$), inextensible, unshearable bending deformations~\cite{audoly2010};  
the strains remain small but with possibly large centerline displacements. The beam is assumed to be in quasi-static moment balance, since the inertia of each of its elements is negligible in the limit $h \ll L$~\cite{liu2021}. 

Under the above assumptions, we represent the deformed centerline by $\br(s,t) = x(s,t)\be_x + y(s,t)\be_y$, where the arclength $s {\in} (0,\,L)$ is measured from the beam's inner end (Fig.~\ref{fig1}e). The tangent angle of the centerline, $\theta(s,t)$, is defined by $\br' = \cos\theta\,\be_x+\sin\theta\,\be_y$ (Fig.~\ref{fig1}f), where we use $(\cdot)'\equiv\partial(\cdot)/\partial s$ and $\dot{(\cdot)}\equiv\partial(\cdot)/\partial t$. The centrifugal, Euler, and Coriolis forces (per unit length) experienced by the beam are, respectively, $\bfo_{\omega} {=} \rho A \Omega^2\br$, $\bfo_e {=} -\rho A \dot{\Omega}\be_z{\times}\br$, and $\bfo_\mathrm{c} {=} -2\rho A\Omega\be_z{\times}\left(\dot{\br}\right)_r$, where $A {=}bh$ is the cross-section area and $(\dot{\br})_r {=} \dot{x}\be_x + \dot{y}\be_y$ is the linear velocity in the rotating FoR; $(\cdot)_r$ denotes differentiation with respect to this frame. Writing $\bn(s,t)$ for the resultant force, the dynamic \emph{elastica} equations, expressing conservation of linear and angular momentum with a linearly elastic (Euler-Bernoulli) constitutive law, are~\cite{audoly2010}:
\begin{align}
     \bn'+\bfo_{\omega}+\bfo_e+\bfo_\mathrm{c} &= \rho A\left(\ddot{\br}\right)_r+\eta\left(\dot{\br}\right)_r, \label{eqn:elasticaforcedim} \\
      B\theta''\be_z+\br'\times\bn &= \mathbf{0}, 
    \label{eqn:elasticamomentdim}
\end{align}
where $B {=} EI$ is the bending modulus, $I {=} h^3b/12$ is the area moment of inertia, and we assume isotropic viscous damping (coefficient $\eta$), which lumps external and material effects. The BCs at the outer end are $\br(L,t) {=} R\be_x$ and $\theta(L,t) {=} 0$ (Figs.~\ref{fig1}d,\,f). At the inner end, we impose $\bn(0,t) {=} \mathbf{0}$, $\theta'(0,t) {=} \kappa_0$ (cantilever), or $\br(0,t) {=} \left[R-L\left(1-\chi\right)\right]\be_x$, $\theta(0,t) {=} 0$ (arch). The unloaded configurations set the initial conditions. 

We proceed by estimating the relative importance of the different underlying forces (per unit length), noting that the centrifugal force scales as $|\bfo_{\omega}| {\sim} \rho A \Omega^2 R$ and the Euler force as $|\bfo_e| {\sim} \rho A \dot{\Omega} R$ (using $|\br|{\sim} R$). Comparing these two with the typical bending force, $|\bn'| {\sim} B/L^3$, yields the dimensionless quantities:
\begin{equation}
    \cC = \frac{\rho A \Omega^2 R L^3}{B} \quad \text{and} \quad \cE = \frac{\rho A \dot{\Omega} R L^3}{B},\label{eqn:defnC,E}
\end{equation}
which we term the centrifugal and Euler numbers. Similar parameters have been identified in related problems~\cite{nowinski1964,tang_note_1970,mostaghel_buckling_1973,chen2007,chen2011,delapierre_wrinkling_2018,sader_shear-induced_2019}, albeit with other geometric factors. The timescale of bending motions is $t^* {=} (\rho A L^4/B)^{1/2}$, obtained from the balance of inertial and bending forces in Eq.~\eqref{eqn:elasticaforcedim}. Thus, assuming the beam deforms by a distance $L$ over $t^*$, the beam velocity scales as $|(\dot{\br})_r| {\sim} L/t^{*}$ and the Coriolis force as $|\bfo_\mathrm{c}| {\sim} \rho A \Omega L/t^*$. Using Eq.~\eqref{eqn:defnC,E}, we find the ratio $|\bfo_\mathrm{c}|L^3/B {=} (\delta\cC)^{1/2}$, where $\delta {=} L/R$. Because, in general, $\cC {=} O(1)$ and $\delta {\ll} 1$ in our experiments, the Coriolis force is negligible.

We non-dimensionalize~\cite{endnoteSuppInfo} Eqs.~\eqref{eqn:elasticaforcedim}--\eqref{eqn:elasticamomentdim} and BCs using the various scales discussed above. The loading is imposed via the time-dependent centrifugal and Euler numbers defined in Eq.~\eqref{eqn:defnC,E}, evaluated using analytical approximations of the experimental velocity/acceleration profiles~\cite{endnoteSuppInfo}. The other dimensionless parameters fixed in each experiment are the geometric ratio $\delta = L/R$ and the dimensionless damping coefficient, $\Upsilon = \eta L^4/(B t^{*})$. We solve the dimensionless equations numerically using the method of lines, following previous work~\cite{gomezthesis,liu2021}: we discretize the equations in space and integrate the resulting set of ordinary differential equations in time~\cite{endnoteSuppInfo}. 

We also perform a weakly-nonlinear analysis of the equilibrium solutions for cantilevered beams~\cite{endnoteSuppInfo}. When $\cE {=} 0$ and $\kappa_0{=}0$, this analysis confirms that the buckling instability is a supercritical pitchfork bifurcation at a critical centrifugal number, $\cC^{\star}(\delta)$. In the limit of $\delta{\ll} 1$ applicable to our experimental system, the linearized problem is equivalent to that of gravitational buckling~\cite{greenhill1881,wang1986}, yielding $\cC^{\star}(0) {\approx} 7.84$. Beyond the buckling onset (\textit{i.e.}, $\cC {>} \cC^{\star}$), $\theta$ is no longer small, requiring numerical solutions. Fig.~\ref{fig2}(b) shows the computed post-buckled equilibrium branch (dashed line), in excellent agreement with the experiments for small but finite accelerations, serving as validation of the model.

As observed in the experimental results presented in Fig.~\ref{fig2}(b), the buckling onset and the ensuing deformation are typically smooth in the presence of imperfections when compared to a perfect pitchfork~\cite{strogatz2015}. Thus, we introduce an empirical definition for the critical centrifugal number, $\cC_\mathrm{c}$, as the $\cC$ value at which the normalized midpoint displacement first exceeds $|Y_\mathrm{m}|{=}0.15$ (stars in Figs.~\ref{fig2}b--d). To further examine the relative importance of the different dynamic effects that are present, we probe the model and decouple the effects resulting from (I) a time-dependent angular velocity and (II) a non-zero Euler force. To do so, we either artificially omit the Euler force while keeping a time-dependent angular velocity or, instead, we ignore time-dependence by varying $\cC$ quasi-statically while maintaining the Euler force. As evidenced by the data in Fig.~\ref{fig2}(d), where we plot  $\cC_\mathrm{c}$ versus $\cE$, the full and quasi-static simulations (squares and triangles, respectively) are in excellent agreement with the experiments (stars). By contrast, the simulations with only time-dependence of the angular velocity (circles) deviate significantly from the experiments (see inset). These results further evidence that unsteady effects are negligible compared to the `imperfection' introduced by the Euler force, \emph{i.e.}, (II) is the relevant scenario.

\begin{figure}[t]
  \centering
  \includegraphics[width=\linewidth]{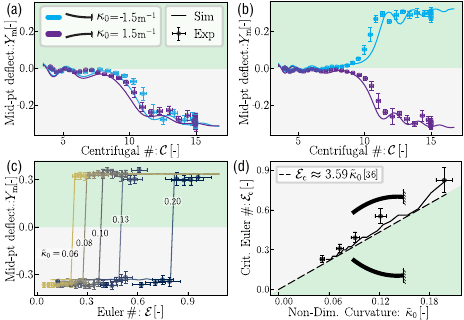}
  \caption{Buckling of naturally curved, cantilevered beams. The legend in (a) applies to all panels. (a,\,b) Normalized midpoint deflection, $Y_\mathrm{m} {=} y_\mathrm{m}/L$, as the centrifugal number, $\cC$, during ramping with Euler number (a) $\cE{=}0.3$ and (b) $\cE{=}0.1$. Each panel shows data for two beams with curvatures $\kappa_0{=}\pm1.5\:\mathrm{m^{-1}}$ (purple/blue lines; see legend). See also S.M. Video 2~\cite{endnoteSuppInfo}. (c) Post-buckled midpoint deflection, $Y_\mathrm{m}(\cC = 15)$, vs.~$\cE$ for beams with normalized natural curvatures $\tilde{\kappa}_0{=}\kappa_0/L {\in} \lbrace0.06,0.08,0.1,0.13,0.2\rbrace$. (d) Critical Euler number $\cE_{\mathrm{c}}$ (at which the buckling direction is inverted) vs.~$\tilde{\kappa}_0$. Also shown is the predicted boundary, Eq.~(S16), from the weakly-nonlinear stability analysis (dashed line)~\cite{endnoteSuppInfo}. }
  \label{fig3}
\end{figure}

While for a straight cantilever, the centrifugal load (along $+\be_x$) always buckles the beam in the Euler-force direction ($-\be_y$), it may be desirable in applications to pre-select the buckling direction in the opposite direction ($+\be_y$).
This can be achieved by fabricating beams with non-zero natural curvature, $\kappa_0$. Fig.~\ref{fig3}(a,\,b) presents results for two such beams possessing equal and opposite natural curvatures, $\kappa_0=\pm1.5\:\mathrm{m^{-1}}$, while fixing all other parameters (experimentally, a single beam can simply be flipped about $\be_x$ before clamping). For large accelerations ($\alpha=10\:\mathrm{rad}/\mathrm{s}^{2}$ in Fig.~\ref{fig3}a; see also panels a--b in S.M. Video 2~\cite{endnoteSuppInfo}), the two cases are nearly identical, with excellent agreement between experiments and simulations. However, for lower accelerations ($\alpha=2.0\:\mathrm{rad}/\mathrm{s}^{2}$ in Fig.~\ref{fig3}b; S.M. Video 2 panels c--d~\cite{endnoteSuppInfo}), the two beams buckle in opposite directions, indicating the geometrical imperfection dominates the transverse Euler force. 

To delineate the transition from a curvature-controlled to an (Euler) force-controlled buckling direction, we consider the beam's midpoint displacement, $Y_\mathrm{m}$, at a fixed centrifugal number beyond buckling ($\cC {=} 15$): in Fig.~\ref{fig3}(c), this is plotted versus $\cE$ for five beams with different normalized natural curvature $\tilde{\kappa}_0{=}\kappa_0/L {\in} [0.06,\,0.2]$. As $\tilde{\kappa}_0$ is increased (\textit{i.e.}, increasingly imperfect beams), the transition from buckling towards $-\be_y$ to $+\be_y$ occurs at higher values of $\cE$, both in experiments (symbols) and simulations (curves). In Fig.~\ref{fig3}(d), we plot the critical Euler number, $\mathcal{E}_{\mathrm{c}}$, for this transition as a function of $\tilde{\kappa}_0$, effectively constructing a phase diagram of the beam's buckling direction. Again, there is excellent agreement between  experiments (symbols), simulations (solid curve), and, notably, also with the boundary $\cE \approx 3.59\:\tilde{\kappa}_0$ (dashed line) predicted from the stability analysis~\cite{endnoteSuppInfo}. While centrifugal forces drive the buckling instability of cantilevered beams, for a given $\tilde{\kappa}_0$ the buckling direction can therefore be selected on-demand via accurate control of $\cE$ according to the phase diagram in Fig.~\ref{fig3}(d). 

\begin{figure}[t]
  \centering
  \includegraphics[width=\linewidth]{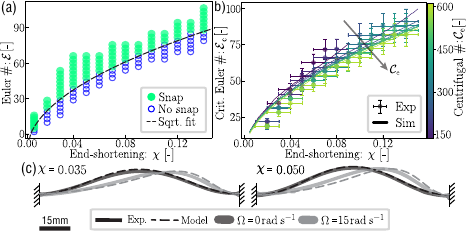}
  \caption{Snap-through of a pre-arched (double-clamped) beam. (a) Phase diagram for the presence (closed symbols) or absence (open symbols) of snapping in the parameter space of Euler number and end-to-end shortening, $(\cE,\,\chi)$ (here $\cC_{\mathrm{e}}{=}373$). (b) Critical Euler number for snapping, $\cE_{\mathrm{c}}$, vs.~$\chi$ at different centrifugal numbers $\cC_{\mathrm{e}}{\in} \lbrace 151,\, 198,\,250,\,309,\,373,\,445,\,522,\,605 \rbrace$. (c) Beam profiles obtained experimentally and numerically (solid and dashed curves, respectively) for $\chi = 0.035$ (left panel) and $\chi = 0.050$ (right). Shapes are shown at rest (dark gray) and rotating at $\Omega{=}15\:\mathrm{rad}/\mathrm{s}$ (light gray). See also S.M. Video 3~\cite{endnoteSuppInfo}}
  \label{fig4}
\end{figure}

Thus far, for cantilevered beams, we showed that the centrifugal force drives buckling while the Euler force lowers the instability onset. We now turn to the pre-arched beams (Fig.~\ref{fig1}f), for which the centrifugal and Euler forces switch roles: the latter drives snapping while the former modulates the instability. In Fig.~\ref{fig4}(a), we present a phase diagram for the presence/absence (closed/open symbols) of snapping in the $(\cE,\,\chi)$ parameter space;  we fix the final centrifugal number at $\cC_{\mathrm{e}}{=}373$ ($\Omega_\mathrm{e}{=}12\:\mathrm{rad}/\mathrm{s}$). Note that the phase boundary, above which snapping occurs, is consistent with the scaling $\cE{\sim}\chi^{1/2}$ (dashed curve); as in other snap-through problems~\cite{gomez2017passive}, this scaling can be rationalized by comparing the typical midpoint deflection (here due to Euler forces) to the initial height of the arch shape. We have thus demonstrated the possibility of actuating rotating mechanisms via the Euler force, whose acting direction can be selected. Euler-actuated mechanisms may switch reversely between stable states, unlike if actuated alone by centrifugal forces, which always act radially outwards.

Although the centrifugal force does not drive snapping, it affects the behavior up to its onset. In Fig.~\ref{fig4}(b) we explore the effect of the centrifugal load, $\cC_{\mathrm{e}}$, on the critical Euler loading for snapping, $\cE_{\mathrm{c}}$, plotted as a function of end-to-end shortening, $\chi$. Even if modest, there is an overall reduction of $\cE_{\mathrm{c}}$ with $\cC_{\mathrm{e}}$, especially for higher values of $\chi$. Qualitatively speaking, when the arched beam is driven at higher angular velocities, its shape becomes increasingly asymmetric (Fig.~\ref{fig4}c) due to the centrifugal force; these asymmetric shapes, of higher bending energy, then lower the energetic barrier for snapping, leading to a lower $\cE_{\mathrm{c}}$. Overall, these results demonstrate that accurate control of the angular velocity can both trigger and modify the snapping of a pre-arched beam.

In conclusion, we demonstrated the feasibility of leveraging both centrifugal and Euler forces to precisely trigger instabilities in rotating \textit{elastica} structures, by accurately controlling their angular velocity drive. Looking ahead, our investigation could be broadened to include more complex loading profiles (\textit{e.g.}, harmonic velocities), diverse geometries (\textit{e.g.}, tapered beams, plates, and shells), and varied material properties (including nonlinear and metamaterial behaviors). These advances have the potential to pave the way for a new class of `programmable' mechanisms that harness the rich instabilities inherent to unsteadily rotating structures, a novel conceptional framework that we refer to as \textit{gyrophilia}. \clearpage

\nocite{ebert2003texturing,ruhoff1996,howell2009,keener1988}

\newpage
\widetext

 \begin{center}
\bf \large \textit{Gyrophilia}: Harnessing Centrifugal and Euler Forces for \\ Tunable Buckling of a Rotating \textit{Elastica}\\ 
\vspace{2mm}
\centerline{\bf \large -- Supplemental Material -- }
\vspace{2mm}
\centerline{\large Eduardo Gutierrez-Prieto, Michael Gomez, and Pedro M. Reis}

\end{center}
\vspace{0mm}


\vspace{0.5cm}
In this document, we provide Supplemental Material (S.M.) for the experimental methods and theoretical calculations presented in the main text. \S\ref{sec:SM_loading} presents some additional details on the experimental loading protocol. In \S\ref{sec:SM_theoryforms}, we detail the \textit{elastica} model that we use to perform dynamic simulations; in particular, we discuss its non-dimensionalization and implementation, including numerical methods. In \S\ref{sec:SM_weaklynonlin}, focusing on cantilevered beams, we present the weakly-nonlinear buckling analysis, which allows us to determine the type of bifurcation underlying the instability as well as the buckling direction. Finally, in \S\ref{sec:SM_videos}, we present the captions for the S.M. Videos $1$--$3$.

\section{Experimental protocol for the rotational loading}
\label{sec:SM_loading}

The rotational loading is set by imposing a time-dependent angular velocity, $\Omega(t)$, in two successive stages. First, in a pre-loading stage, $\Omega(t)$ is slowly ramped from zero to $\Omega_0 > 0$ at an acceleration $\dot{\Omega} = 5\:\mathrm{rad}\,\mathrm{s}^{-2}$, before being held at $\Omega_0$ for $2\:\mathrm{s}$, ensuring the decay of any transient oscillations. Next, in the second loading stage, $\Omega(t)$ is ramped from $\Omega_0$ to $\Omega_e$ (where $\Omega_e > \Omega_0$); throughout this stage, the angular acceleration is constant, i.e., $\dot{\Omega} = \alpha = \mathrm{constant}$, except for short `jerk intervals' (duration $t_j = 100\:\mathrm{ms}$) at the start/end of the ramping when $\dot{\Omega}$ rapidly varies between $\alpha$ and zero. Two representative time series of $\Omega(t)$ and $\dot{\Omega}(t)$ are presented in Fig.~1(b) and Fig.~1(c) of the main text, respectively.

\section{Theoretical formulations}
\label{sec:SM_theoryforms}

\subsection{Non-dimensionalization of the dynamic \textit{elastica} equations}

The dynamic \textit{elastica} equations were presented Eqs.~(1)--(2) of the main text. To non-dimensionalize these equations, it is natural to scale lengths by the beam length, $L$, time by the inertia-bending timescale, $t^* = (\rho A L^4/B)^{1/2}$, and forces by the typical buckling force, $B/L^2$. Denoting dimensional and dimensionless quantities by lowercase and uppercase variables, respectively, we write:
\begin{equation*}
    s = LS, \quad t = t^* T, \quad (x,y) = (R-L+LX,LY), \quad (n_x,n_y) = \frac{B}{L^2}(N_X,N_Y),
\end{equation*}
where we have expanded $\bn = n_x\be_x+n_y\be_y$. Note that we have included a shift of $R-L$ in the definition of $X$, so that $X=1$ at the outer end of the beam. Substituting the expressions for the fictitious forces $(\bfo_{\omega},\bfo_{e},\bfo_{c})$, and applying the above re-scalings, Eqs.~(1)--(2) and the geometric relation $\br' = \cos\theta\,\be_x+\sin\theta\,\be_y$ can be written in component form as (dropping the shorthand $(\cdot)'$ and $\dot{(\cdot)}$ to avoid confusing dimensional/dimensionless derivatives)
\begin{align}
& \pd{N_X}{S}+\cC\left(1-\delta+\delta X\right)+\delta\cE Y+2(\delta\cC)^{1/2}\pd{Y}{T} = \pdd{X}{T}+\Upsilon\pd{X}{T}, \nonumber \\
& \pd{N_Y}{S}+\delta\cC Y -\cE\left(1-\delta+\delta X\right)-2(\delta\cC)^{1/2}\pd{X}{T} = \pdd{Y}{T}+\Upsilon\pd{Y}{T}, \nonumber \\
& \pdd{\theta}{S}-N_X\sin\theta+N_Y\cos\theta = 0, \qquad \pd{X}{S} = \cos\theta, \quad \pd{Y}{S} = \sin\theta.
\label{eqn:elasticasystem}
\end{align}
Here $\cC$ and $\cE$ are, respectively, the dimensionless centrifugal and Euler numbers defined in Eq.~(3) of the main text. The other dimensionless parameters are $\delta = L/R$ and $\Upsilon = \eta L^4/(Bt^*)$. We note that $\delta\geq 0$ compares the beam's size, $L$, to the radial distance, $R$, and hence measures the relative variation of the fictitious forces along the beam's length. 

Recalling the discussion below Eq.~(2) of the main text, the dimensionless boundary conditions (BCs) are:
\begin{equation}
X(1,T) = 1, \quad Y(1,T) = 0, \quad \theta(1,T) = 0 \quad \mathrm{and} \quad 
\begin{cases} N_X(0,T) = 0, \quad N_Y(0,T) = 0, \quad \pd{\theta}{S}(0,T) = \hat{\kappa}_0 \quad \mathrm{(cantilever)}; \\
\ \ \ X(0,t) = \chi, \ \ \quad Y(0,T) = 0, \ \ \quad \theta(0,T) = 0 \ \ \quad \mathrm{(arch)},
 \end{cases}  \label{eqn:elasticaBCs}
\end{equation}
where $\hat{\kappa}_0 = L\kappa_0$ is the dimensionless natural curvature. The system is closed by the initial conditions discussed below.

\subsection{Value of the damping coefficient, $\eta$}
\label{SM_sec:dampingcoeff}

In our model, the damping coefficient $\eta$ was measured from underdamped oscillations of the beams (in a cantilevered configuration) in the absence of rotational loading. According to linear stability analysis~[48], small-amplitude underdamped oscillations decay in time like $e^{-t/[t]_d}$, where the time constant is $[t]_d = 2\rho A/\eta$. Experimentally, we perturbed the beams and extracted the time series of the tip displacement; the time constant $[t]_d$ was then determined by fitting the envelope of the oscillations. Using the known values of $\rho$ and $A$ (see main text), we obtain $\eta = (0.0323\pm 0.003)\:\mathrm{Pa}.\mathrm{s}$ and $\eta = (0.0427\pm 0.003)\:\mathrm{Pa}.\mathrm{s}$ for the beams made of VPS32 and VPS22, respectively. 

\subsection{Rotational loading model}

In our dimensionless model,  Eqs.~(\ref{eqn:elasticasystem})--(\ref{eqn:elasticaBCs}), the rotational loading is imposed via time-dependent centrifugal and Euler numbers evaluated via Eq.~(3), using analytical expressions for $\Omega(t)$ and $\dot{\Omega}(t)$ that approximate the angular velocity and acceleration imposed experimentally. In particular, we simulate the two dynamic stages (pre-loading and loading) described in \S\ref{sec:SM_loading}. During the second (loading) stage, we approximate the angular acceleration during each jerk interval using the so-called \textit{smootherstep} function~[50], denoted $S_2$. For a general quantity $a(t)$, this function is a monotonic ramp between the points $(a,t) = (a_0,t_0)$ and $(a_1,t_1)$, with zero first and second-order derivatives at the end-points:
\begin{equation*}
    S_2(t;t_0,t_1,a_0,a_1) = a_0 +(a_1-a_0)\left[6\left(\frac{t-t_0}{t_1-t_0}\right)^5-15\left(\frac{t-t_0}{t_1-t_0}\right)^4+10\left(\frac{t-t_0}{t_1-t_0}\right)^3\right] \qquad t\in(t_0,t_1).
\end{equation*}
With $\Delta t$ denoting the duration of the loading stage (taken to start at $t = 0$), the angular acceleration during the loading stage is modeled as the piece-wise function
\begin{equation*}
    \dot{\Omega}(t) =
    \begin{cases}
    \begin{aligned}
        S_2(t;0,t_j,0,\alpha) \qquad & \ \  \qquad 0 \leq t < t_j, \\
        \alpha \qquad & \ \qquad t_j \leq t < \Delta t-t_j, \\
        S_2(t;\Delta t-t_j,\Delta t,\alpha,0) \qquad &\Delta t-t_j \leq t < \Delta t. \\
        \end{aligned}
    \end{cases}
\end{equation*}
The corresponding angular velocity, $\Omega(t)$, can be determined by integration using the initial condition $\Omega(0) = \Omega_0$. The angular velocity at the end of the loading stage is then  $\Omega(\Delta t) = \Omega_0 + \alpha(\Delta t - t_j)$. Thus, to satisfy the imposed final value $\Omega(\Delta t) = \Omega_e$, we choose $\Delta t = \left(\Omega_e-\Omega_0\right)/\alpha+t_j$.

\subsection{Initial conditions}

At the start of the pre-loading stage, the beam is at rest and in equilibrium (absence of external loads). For cantilevered beams, the beam is stress-free in this equilibrium and everywhere adopts its natural curvature $\hat{\kappa}_0$:
\begin{equation*}
    N_X(S) = 0, \quad N_Y(S) = 0, \quad \theta(S) = -\hat{\kappa}_0(1-S), \quad X(S) = 1-\frac{\sin\left[\hat{\kappa}_0\left(1-S\right)\right]}{\hat{\kappa}_0}, \quad Y(S) = \frac{1-\cos\left[\hat{\kappa}_0\left(1-S\right)\right]}{\hat{\kappa}_0}.
\end{equation*}
This may be verified as the equilibrium solution of Eqs.~\eqref{eqn:elasticasystem}--\eqref{eqn:elasticaBCs} when $\cC = \cE = 0$.

In the case of pre-arched beams (with $\hat{\kappa}_0=0$), the beam undergoes Euler (pre-)buckling in the absence of external loads due to the imposed end-to-end shortening, $\chi$, between the double-clamped boundaries. To determine the beam shape, we solve numerically the steady version of Eqs.~\eqref{eqn:elasticasystem}--\eqref{eqn:elasticaBCs} (with $\cC = \cE = 0$). This is achieved by discretizing the equations (in an identical manner to our dynamic simulations; see \S\ref{sec:SM_numerics}) and solving the resulting set of algebraic equations in MATLAB using the routine \texttt{fsolve}. As an initial guess, we use the linearized solution for $\theta \ll 1$ corresponding to mode-1 Euler buckling, i.e., $N_X(S) = -4\pi^2$, $N_Y(S) = 0$, $\theta(S) = 2\chi^{1/2}\sin\left(2\pi S\right)$, $X(S) = S$ and $Y(S) = \chi^{1/2}\left[1-\cos\left(2\pi S\right)\right]/\pi$.

\subsection{Numerical solution}
\label{sec:SM_numerics}

To solve the Eqs.~\eqref{eqn:elasticasystem}--\eqref{eqn:elasticaBCs} numerically, we use the method of lines~[47]. We discretize the interval $S\in[0,1]$ using a uniform mesh with spacing $\Delta S = 1/N$ (there are $N+1$ grid points in total), resulting in a system of ordinary differential equations (ODEs) in time. Before discretizing, it is advantageous to recast Eq.~\eqref{eqn:elasticasystem} as a single integro-differential equation involving the tangent angle $\theta(S,T)$ only: this means we do not need to explicitly impose inextensibility of the beam's centerline between each grid point, allowing us to avoid a large number of constraints and, thus,  integrating the equations using pre-existing ODE solvers efficiently.

To this end, we first integrate the geometric relations in Eq.~\eqref{eqn:elasticasystem} with the BCs $X(1,T) = 1$, $Y(1,T) = 0$ from Eq.~\eqref{eqn:elasticaBCs} to obtain $X(S,T) = 1-\int_S^{1}\cos\theta(\eta,T)\,\id\eta$ and $Y(S,T) = -\int_S^{1}\sin\theta(\eta,T)\,\id\eta$. Substituting the above expressions into the force balances in Eq.~\eqref{eqn:elasticasystem}, and integrating again with BCs $N_X(0,T) = P(T)$, $N_Y(0,T) = Q(T)$ (we do not restrict to the cantilever/pre-arched scenarios for now), we can determine the force resultants as
\begin{align*}
    N_X(S,T) & = P(T)-\cC S + \int_0^S\int_\xi^1\left\lbrace\left[\pdd{\theta}{T}+\Upsilon\pd{\theta}{T}+\delta\cE\right]\bigg\lvert_{S=\eta}\sin\theta(\eta,T)+\left[\pd{\theta}{T}+\left(\delta\cC\right)^{1/2}\right]^2\bigg\lvert_{S=\eta}\cos\theta(\eta,T)\right\rbrace\id\eta\,\id\xi, \\
    N_Y(S,T) & = Q(T)+\cE S -\int_0^S\int_\xi^1\left\lbrace\left[\pdd{\theta}{T}+\Upsilon\pd{\theta}{T}+\delta\cE\right]\bigg\lvert_{S=\eta}\cos\theta(\eta,T)-\left[\pd{\theta}{T}+\left(\delta\cC\right)^{1/2}\right]^2\bigg\lvert_{S=\eta}\sin\theta(\eta,T)\right\rbrace\id\eta\,\id\xi.
\end{align*}
Inserting these expressions into the moment balance in Eq.~\eqref{eqn:elasticasystem}, and making use of the addition formulae for $\sin\left[\theta(S,T)-\theta(\eta,T)\right]$ and $\cos\left[\theta(S,T)-\theta(\eta,T)\right]$, we arrive at
\begin{align}
    0  = & \pdd{\theta}{S}-\left(P-\cC S\right)    \sin\theta+\left(Q+\cE S\right)\cos\theta \nonumber \\
    & \: -\int_0^S\int_\xi^1\left\lbrace\left[\pdd{\theta}{T}+\Upsilon\pd{\theta}{T}+\delta\cE\right]\bigg\lvert_{S=\eta}\cos\left[\theta(S,T)-\theta(\eta,T)\right]+\left[\pd{\theta}{T}+\left(\delta\cC\right)^{1/2}\right]^2\bigg\lvert_{S=\eta}\sin\left[\theta(S,T)-\theta(\eta,T)\right]\right\rbrace\id\eta\,\id\xi. \label{eqn:elasticathetaonly}
\end{align}
The remaining BCs in \eqref{eqn:elasticaBCs} in terms of $\theta$ become
\begin{equation}
 \begin{cases}
     P = Q = 0, \quad  \pd{\theta}{S}(0,T) = \hat{\kappa}_0, \quad \theta(1,T) = 0 \quad \mathrm{(cantilever)}; \\
     \theta(0,T) = 0, \quad \theta(1,T) = 0, \quad \int_0^1\cos\theta(S,T)\,\id S = 1-\chi,\quad \int_0^1\sin\theta(S,T)\,\id S = 0 \quad \mathrm{(arch)}.
     \end{cases}
\label{eqn:elasticaBCsthetaonly}
\end{equation}
For arched beams, $P$ and $Q$ are unknown and act as Lagrange multipliers to enforce the integral constraints in \eqref{eqn:elasticaBCsthetaonly}.

Our discretization of Eqs.~\eqref{eqn:elasticathetaonly}--\eqref{eqn:elasticaBCsthetaonly}  follows that employed previously by Refs.~[46,47] to simulate the snap-through dynamics of an \textit{elastica}. We formulate a scheme with second-order accuracy as $\Delta S\to 0$: we approximate the $\partial^2\theta/\partial S^2$ term in Eq.~\eqref{eqn:elasticathetaonly} using a second-order centered difference on the numerical mesh, and we use the trapezium rule to approximate integrals. The resulting system of ODEs are written in matrix-vector form and integrated in MATLAB using the solver \texttt{ode15s}. For pre-arched beams, the integral constraints in Eq.~\eqref{eqn:elasticaBCsthetaonly} mean that the system is differential-algebraic (since the Lagrange multipliers $P$ and $Q$ do not explicitly enter Eq.~\eqref{eqn:elasticaBCsthetaonly}). This can be avoided using the method described in Ref.~[51]: we differentiate the integral constraints twice in time, then eliminate $\partial^2\theta/\partial T^2$ terms (using the discretized form of Eq.~\eqref{eqn:elasticathetaonly}) to obtain a closed linear system for $P$ and $Q$. In all simulations reported in the main text, we take $N = 100$;  having checked that the results are insensitive to increasing $N$ or decreasing integration tolerances. Each simulation typically completes in a few seconds on a laptop computer.

\section{Weakly-nonlinear analysis of buckling of cantilevered beams}
\label{sec:SM_weaklynonlin}

In this section, focusing on cantilevered beams, we perform a weakly-nonlinear analysis of the equilibrium solutions near the buckling onset, similar to that performed in other buckling problems~[52]. We assume a small Euler number, $\cE\ll 1$, and small natural curvature, $\hat{\kappa}_0\ll 1$, so that the amplitude of the solution before buckling is small: $\theta \ll 1$. We write $C^{\star}$ for the value of the centrifugal number $\cC$ at the buckling onset (to be determined). We then perturb
\begin{equation}
    \cC = \cC^{\star} + \epsilon\,\cC^{(1)}, \quad \theta = \epsilon^{1/2}\left(\theta^{(0)}+\epsilon\,\theta^{(1)}+\ldots\right),
    \label{eqn:expandC,theta}
\end{equation}
where $\epsilon \ll 1$ is a fixed parameter and $\cC^{(1)}$ is a control parameter with $\cC^{(1)} > 0$, $\cC^{(1)} = O(1)$; the asymptotic expansion of $\theta$ begins at $O(\epsilon^{1/2})$ because we anticipate a supercritical pitchfork bifurcation, in which the amplitude of the solution grows like the \emph{square root} of the perturbation beyond buckling~[52].

To proceed, it is convenient to use the integro-differential formulation of the \textit{elastica} equations derived in \S\ref{sec:SM_numerics}. Neglecting time derivatives (and setting $P = Q = 0$), Eq.~\eqref{eqn:elasticathetaonly} becomes
\begin{equation}
    0 = \dd{\theta}{S}+\cC S\sin\theta+\cE S\cos\theta -\delta\int_0^S\int_\xi^1\bigl(\cE\cos\left[\theta(S)-\theta(\eta)\right]+\cC\sin\left[\theta(S)-\theta(\eta)\right]\bigr)\id\eta\,\id\xi, \label{eqn:elasticathetaonlysteadycantilever}
\end{equation}
to be solved with the BCs~\eqref{eqn:elasticaBCsthetaonly}. We substitute the asymptotic expansions \eqref{eqn:expandC,theta} into Eqs.~\eqref{eqn:elasticathetaonlysteadycantilever} and \eqref{eqn:elasticaBCsthetaonly}, and solve at successive orders in $\epsilon$. In what follows, we assume that $\cE = O(\epsilon^{3/2})$ and $\hat{\kappa}_0 = O(\epsilon^{3/2})$, so that the intrinsic curvature and Euler number only enter the problem at first order, i.e., at $O(\epsilon^{3/2})$. In practice, given $\cE \ll 1$ and $\hat{\kappa}_0 \ll 1$, this can be satisfied by choosing $\left[\mathrm{max}\left(\cE,\hat{\kappa}_0\right)\right]^{2/3} \lesssim \epsilon \ll 1$.

\subsection{Leading-order problem: $O(\epsilon^{1/2})$}

At $O(\epsilon^{1/2})$, Eq.~\eqref{eqn:elasticathetaonlysteadycantilever} and BCs~\eqref{eqn:elasticaBCsthetaonly} yield the linear eigenvalue problem:
\begin{equation}
    \mathcal{L}\theta^{(0)} = 0, \quad \d{\theta^{(0)}}{S}(0) = 0, \quad \theta^{(0)}(1) = 0 \qquad \mathrm{where} \qquad \mathcal{L}\theta \equiv \dd{\theta}{S}+\cC^{\star} S\theta-\delta\cC^{\star}\int_0^S\int_\xi^1\left[\theta(S)-\theta(\eta)\right]\id\eta\,\id\xi. \label{eqn:leadingorderproblem}
\end{equation}
We can write $\theta^{(0)} = A^{(0)}\phi^{(0)}$, where $A^{(0)}$ is an unknown (scalar) amplitude and $\phi^{(0)}$ satisfies the eigenvalue problem with normalization condition $\phi^{(0)}(0) =1$, i.e., 
\begin{equation}
    \mathcal{L}\phi^{(0)} = 0, \quad \phi^{(0)}(0) = 1, \quad \d{\phi^{(0)}}{S}(0) = 0, \quad \phi^{(0)}(1) = 0.
    \label{eqn:phiproblem}
\end{equation}\\

\noindent \textbf{\textit{Simplification as $\delta \to 0$:}}
In the limit $\delta \to 0$ (i.e., $L/R\to 0$), the spatial variation of the fictitious forces along the beam length is negligible. The centrifugal force acts analogously to a uniform gravitational field so that the leading-order problem (which only involves the centrifugal force) is equivalent to the classical Greenhill problem for gravitational buckling~[40,42]]. The equation $\mathcal{L}\phi^{(0)} = 0$ reduces to the Airy equation $\id^2\phi^{(0)}/\id S^2 +\cC^{\star} S\phi^{(0)} = 0$; the solution satisfying the BCs at $S = 0$ (see Eq.~\eqref{eqn:phiproblem}) is 
\begin{equation}
    \phi^{(0)}(S) = \frac{3^{1/6}\Gamma(2/3)}{2}\left[\sqrt{3}\,\mathrm{Ai}\left(-{\cC^{\star}}^{1/3}S\right)+\mathrm{Bi}\left(-{\cC^{\star}}^{1/3}S\right)\right]. \label{eqn:phisolnzerodelta}
\end{equation}
The BC $\phi^{(0)}(1) = 0$ then yields  $\sqrt{3}\,\mathrm{Ai}\left(-{\cC^{\star}}^{1/3}\right)+\mathrm{Bi}\left(-{\cC^{\star}}^{1/3}\right) = 0$, the first positive root of which is $\cC^{\star} \approx 7.84$ (as reported in the main text).\\

\begin{figure}[h!]
  \centering
  \includegraphics[width=0.8\textwidth]{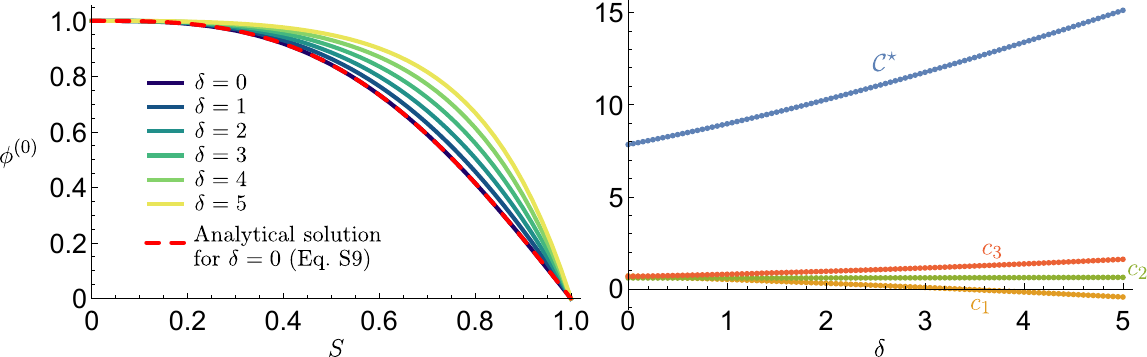}
  \caption{Solution to leading-order problem, Eq.~\eqref{eqn:phiproblem}, obtained by shooting. (Left) Solution for $\phi^{(0)}(S)$ for different values of $\delta$ (see legend). (Right) Corresponding values of the eigenvalue $\cC^{\star}$ and the constants $c_1$, $c_2$ and $c_3$ (defined in Eq.~\eqref{eqn:defnc1c2c3}).}
  \label{SM_fig1}
\end{figure}

\noindent \textbf{\textit{General $\delta \neq 0$:}} For non-zero $\delta$, the integral term in the operator $\mathcal{L}(\cdot)$ means that an analytical solution to $\mathcal{L}\phi^{(0)} = 0$ is generally not possible, so we solve Eq.~\eqref{eqn:phiproblem} numerically using a shooting method: we solve $\mathcal{L}\phi^{(0)} = 0$ with the conditions at $S = 0$ to determine $\phi^{(0)}(S;\cC^{\star})$, then we apply the remaining BC $\phi^{(0)}(1;\cC^{\star}) = 0$ to determine $\cC^{\star}$. The shooting is performed in Mathematica using the routines \texttt{ParametricNDSolve} and \texttt{FindRoot}.  Fig.~\ref{SM_fig1}(left) shows the numerically determined $\phi^{(0)}(S)$ for several values of $\delta\in[0,5]$ (left) and the corresponding values of $\cC^{\star}(\delta)$ (right).

\subsection{First-order problem: $O(\epsilon^{3/2})$}

At $O(\epsilon^{3/2})$, Eq.~\eqref{eqn:elasticathetaonlysteadycantilever} and BCs~\eqref{eqn:elasticaBCsthetaonly} become
\begin{align}
    \mathcal{L}\theta^{(1)} = & -S\left(\epsilon^{-3/2}\cE+\cC^{(1)}\theta^{(0)}-\frac{\cC^{\star}}{6}{\theta^{(0)}}^3\right) \nonumber \\
    & +\delta\int_0^S\int_\xi^1\left\lbrace\epsilon^{-3/2}\cE+\cC^{(1)}\left[\theta^{(0)}(S)-\theta^{(0)}(\eta)\right]-\frac{\cC^{\star}}{6}\left[\theta^{(0)}(S)-\theta^{(0)}(\eta)\right]^3\right\rbrace\id\eta\,\id\xi,  \label{eqn:firstorderproblemfulleqn} \\
    \d{\theta^{(1)}}{S}(0) & = \epsilon^{-3/2}\hat{\kappa}_0, \quad \theta^{(1)}(1) = 0,
    \label{eqn:firstorderproblemBCS}
\end{align}
where $\mathcal{L}(\cdot)$ is the same operator defined in the leading-order problem \eqref{eqn:leadingorderproblem}. After substituting $\theta^{(0)}=A^{(0)}\phi^{(0)}$ into Eq.~\eqref{eqn:firstorderproblemfulleqn}, integrating terms on the right-hand side that are independent of $\eta$, and re-arranging, Eq.~\eqref{eqn:firstorderproblemfulleqn} becomes
\begin{align}
    \mathcal{L}\theta^{(1)} = & -\left[\left(1-\delta\right)S+\frac{\delta}{2}S^2\right]\epsilon^{-3/2}\cE-\cC^{(1)}\left\lbrace\left[\left(1-\delta\right)S+\frac{\delta}{2}S^2\right]\phi^{(0)}-\delta U^{(0)}\right\rbrace A^{(0)} \nonumber \\
    & \: + \frac{\cC^{\star}}{6}\left\lbrace\left[\left(1-\delta\right)S+\frac{\delta}{2}S^2\right]{\phi^{(0)}}^3-3\delta{\phi^{(0)}}^2 U^{(0)}+3\delta\phi^{(0)}V^{(0)}-\delta W^{(0)}\right\rbrace {A^{(0)}}^3, \label{eqn:firstorderproblemeqn}
\end{align}
where we have introduced the variables
\begin{equation*}
    U^{(0)}(S) \equiv -\int_0^S\int_\xi^1\phi^{(0)}(\eta)\,\id\eta\,\id\xi, \quad V^{(0)}(S) \equiv -\int_0^S\int_\xi^1\left[\phi^{(0)}(\eta)\right]^2\id\eta\,\id\xi, \quad W^{(0)}(S) \equiv -\int_0^S\int_\xi^1\left[\phi^{(0)}(\eta)\right]^3\id\eta\,\id\xi. 
\end{equation*}

Because the homogeneous problem $\mathcal{L}(\cdot) = 0$ has the non-trivial solution $\phi^{(0)}$ (the solution of the leading-order problem \eqref{eqn:phiproblem}), the Fredholm Alternative Theorem~[53] implies that a solution for $\theta^{(1)}$ exists only if the right-hand side of Eq.~\eqref{eqn:firstorderproblemeqn} satisfies a solvability condition. This solvability condition can be found by multiplying Eq.~\eqref{eqn:firstorderproblemeqn} by $\phi^{(0)}$ and integrating over $S\in(0,1)$. After integrating by parts, making use of Eq.~\eqref{eqn:phiproblem} and the BCs~\eqref{eqn:firstorderproblemBCS}, the terms in $\theta^{(1)}$ vanish and we are left with an equation for the leading-order amplitude $A^{(0)}$:
\begin{equation}
    0 = \epsilon^{-3/2}\left(\hat{\kappa}_0-c_1\cE\right) - c_2\,\cC^{(1)}A^{(0)}+c_3 {A^{(0)}}^3,
    \label{eqn:amplitudeeqn}
\end{equation}
where we define the constants
\begin{align}
    & c_1 = \int_0^1\left[\left(1-\delta\right)S+\frac{\delta}{2}S^2\right]\phi^{(0)}(S)\,\id S, \quad c_2 = \int_0^1\left\lbrace\left[\left(1-\delta\right)S+\frac{\delta}{2}S^2\right]\left[\phi^{(0)}(S)\right]^2-\delta U^{(0)}(S)\phi^{(0)}(S)\right\rbrace\id S, \nonumber \\
    & c_3 = \frac{\cC^{\star}}{6}\int_0^1\left\lbrace\left[\left(1-\delta\right)S+\frac{\delta}{2}S^2\right]\left[\phi^{(0)}(S)\right]^4-3\delta U^{(0)}(S)\left[\phi^{(0)}(S)\right]^3+3\delta V^{(0)}(S)\left[\phi^{(0)}(S)\right]^2-\delta W^{(0)}(S)\phi^{(0)}(S)\right\rbrace\id S. \label{eqn:defnc1c2c3}
\end{align}
For each $\delta$, these constants are evaluated using the numerical solution for $\phi^{(0)}$. \\

\noindent \textbf{\textit{Type of bifurcation at buckling $(\hat{\kappa}_0 = 0,\ \cE = 0)$:}} When $\hat{\kappa}_0 = 0$ and $\cE = 0$,
Eq.~\eqref{eqn:amplitudeeqn} has the form of an amplitude equation associated with a pitchfork bifurcation:
\begin{equation}
A^{(0)}\left(c_2\cC^{(1)}-c_3{A^{(0)}}^2\right) = 0. \label{eqn:amplitudeeqnperfect}
\end{equation}
Using Eq.~\eqref{eqn:phiproblem} and the relation $\id^2 U^{(0)}/\id S^2 = \phi^{(0)}$ (and the fact that $\cC^{\star} > 0$), it may be shown that $c_2 > 0$ for all $\delta \geq 0$. Moreover, while we are unable to show this analytically, plotting the numerically determined $c_3$ as a function of $\delta$ confirms that $c_3 > 0$; see Fig.~\ref{SM_fig1}(right). Consequently, below the buckling threshold ($\cC^{(1)} < 0$), the only real solution to Eq.~\eqref{eqn:amplitudeeqnperfect} is $A^{(0)} = 0$; above the buckling threshold ($\cC^{(1)} > 0$), Eq.~\eqref{eqn:amplitudeeqnperfect} permits the non-zero solutions $A^{(0)} = \pm(c_2\cC^{(1)}/c_3)^{1/2}$. We conclude that the buckling of straight cantilevered beams under quasi-static changes in centrifugal number $\cC$ corresponds to a \emph{supercritical} pitchfork bifurcation.\\ 

\noindent \textbf{\textit{Direction of buckling $(\hat{\kappa}_0,\ \cE \neq 0)$:}}
Equation \eqref{eqn:amplitudeeqn} indicates that the natural curvature, $\hat{\kappa}_0$, and Euler number, $\cE$, behave analogously to symmetry-breaking imperfections. When these quantities are non-zero, the pitchfork bifurcation becomes `unfolded': the amplitude $A^{(0)}$ smoothly varies from zero as $\cC$ is quasi-statically increased past $\cC^{\star}$ (meanwhile, the other buckled solution in the pair $A^{(0)} = \pm(c_2\cC^{(1)}/c_3)^{1/2}$ forms a disconnected branch). We can infer the direction of buckling from the sign of the constant term in Eq.~\eqref{eqn:amplitudeeqn}, namely $\epsilon^{-3/2}\left(\hat{\kappa}_0-c_1\cE\right)$, as this term determines the sign of the amplitude $A^{(0)}$ (since $c_2 > 0$). In particular, when $\hat{\kappa}_0 > 0$ (corresponding to a beam curved towards the positive $y$ direction), the beam buckles in the direction of its natural curvature when $\cE < \cE_c \equiv \hat{\kappa}_0/c_1$; and, instead, buckles in the direction of the Euler force (negative $y$ direction) when $\cE > \cE_c$. While in general $c_1$ must be evaluated numerically, in the simplified case $\delta = 0$, we can obtain an analytical expression using the analytical solution for $\phi^{(0)}$ given above, in Eq.~\eqref{eqn:phisolnzerodelta}. Substituting this solution into Eq.~\eqref{eqn:defnc1c2c3}, and simplifying using standard identities for Airy functions and the fact that $\cC^{\star}$ ($\approx 7.84$) satisfies the eigenvalue equation $\sqrt{3}\,\mathrm{Ai}(-{\cC^{\star}}^{1/3})+\mathrm{Bi}(-{\cC^{\star}}^{1/3}) = 0$, we obtain $c_1 = 3^{-1/3}\Gamma(1/3)^{-1}{\cC^{\star}}^{-2/3}{\mathrm{Ai}(-{\cC^{\star}}^{1/3})}^{-1}$ and hence
\begin{equation}
    \cE_c = 3^{1/3}\Gamma\left(\frac{1}{3}\right){\cC^{\star}}^{2/3}\mathrm{Ai}\left(-{\cC^{\star}}^{1/3}\right)\hat{\kappa}_0 \approx 3.59\,\hat{\kappa}_0 \qquad (\mathrm{change\ of\ buckling\ direction\ when\ }\delta = 0). \label{eqn:bucklingdirection}
\end{equation}
This equation is used to plot the phase boundary (black dashed line) in Fig.~3(d) of the main text. We also note that, while we have assumed $\cE = O(\epsilon^{3/2})$ and $\hat{\kappa}_0 = O(\epsilon^{3/2})$ in the above analysis, the same expression for $\cE_c$ is obtained when the natural curvature and Euler number enter the problem at leading-order, i.e., at $O(\epsilon^{1/2})$ (not shown here).

\subsection{Quasi-static assumption}

\vspace{-0.1cm}
Given an angular acceleration $\dot{\Omega}$, the timescale over which the angular velocity changes appreciably is $\sim \Omega/\dot{\Omega}$. Thus, over a timescale $t \ll \Omega/\dot{\Omega}$, the variation of $\Omega$ is negligible, and it can be considered constant. On the other hand, since the dynamics are underdamped, we expect any oscillations to decay sufficiently when $t \gg [t]_d$ where $[t]_d = 2\rho A/\eta$ is the decay timescale discussed earlier in \S\ref{SM_sec:dampingcoeff}. Combining these two observations, we expect that the beam is in quasi-static equilibrium over a timescale $\rho A/\eta \ll t \ll \Omega/\dot{\Omega}$ (in dimensionless terms, this corresponds to $\Upsilon^{-1}\ll T \ll \delta^{-1/2}\cC^{1/2}\cE^{-1}
$). This observation explains why, for cantilevered beams, our dynamic simulations and experiments are in excellent agreement with the quasi-static predictions for accelerations $\alpha\lesssim 5\:\mathrm{rad}\,\mathrm{s}^{-2}$; see Fig.~2(d) and Fig.~3(d) of the main text. Indeed, using $\Omega \sim 10\,\,\mathrm{rad}\,\mathrm{s}^{-1}$ (the typical value at the buckling onset; see Fig.~2d) and $\dot{\Omega} \sim 5\:\mathrm{rad}\,\mathrm{s}^{-2}$, we have $\Omega/\dot{\Omega} \sim 2\:\mathrm{s}$  while we calculate the decay timescale as $\rho A/\eta \sim 0.67\:\mathrm{s}$. For pre-arched beams, the quasi-static assumption is generally not satisfied due to the much larger accelerations required for snap-through.

\section{Supplemental Videos}
\label{sec:SM_videos}

\noindent \textbf{Video 1 -- Experimental set-up and sample experiment:} The video corresponds to Fig.~1 of the main text, providing a representative example of an experimental run with a cantilevered beam. The main video is imaged in the lab frame, whereas the inset is imaged in the rotating FoR using the camera mounted on the rotating arm. The geometric and material parameters of the naturally straight ($\kappa_0 {=} 0$) 
beam have been specified in the main text. The angular dynamics of rotation drive are characterized by an initial angular velocity $\Omega_0 {=} 5\:\mathrm{rad}/\mathrm{s}$, final angular velocity $\Omega_{\mathrm{e}} = 15 \:\mathrm{rad}/\mathrm{s}$, and maximum angular acceleration $\alpha=10\:\mathrm{rad}/\mathrm{s}^{2}$. 

\vspace{0.2cm}
\noindent \textbf{Video 2 -- Selecting the buckling direction of an intrinsically curved:}
The video corresponds to Fig.~3(a,b) of the main text, providing a view of two cantilevered specimens of natural curvatures ($\kappa_0 {=} \pm 1.5\:\mathrm{m}^{-1}$) in the rotating FoR, as imaged by the camera mounted on the rotating arm. These beams are rotated with two values of acceleration $\alpha = \{1.5,\,10\}\:\mathrm{rad}/\mathrm{s}^{2}$ while fixing the initial angular velocity $\Omega_0 {=} 5\:\mathrm{rad}/\mathrm{s}$ and the final angular velocity $\Omega_\mathrm{e} {=} 15\:\mathrm{rad}/\mathrm{s}$. The two specimens are fabricated with the same geometry but opposite natural curvature; $\kappa_0 
{=} \pm1.5\:\mathrm{m}^{-1}$. The other geometric and material properties of the beam have been specified in the main text.
Note that the reproduction speed of the videos is different for the two values of acceleration; (a,\,b) sped-up factor $\times$1.5 w.r.t. real-time, and (c,\,d) slowed-up factor $\times$0.3 w.r.t. real-time.

\vspace{0.2cm}
\noindent \textbf{Video 3 -- Acceleration-driven snapping of a rotating pre-arched beam:}
The video corresponds to Fig.~4(a,b) of the main text, providing a view of pre-arched beams in the rotating FoR, as imaged by the camera mounted on the rotating arm. These beams are subject to two levels of end-to-end shortening, $\chi {=} \{0.07,0.09\}$, and are rotated with two values of final angular velocity $\Omega_{\mathrm{e}} {=} \{12,\,15\}\:\mathrm{rad}/\mathrm{s}^{1}$ and for different values of acceleration, $\alpha$ (see legend in the video). The specimen is fabricated according to geometric and material properties specified in the main text.

\end{document}